%Paper: hep-th/9306045
%From: dafr@math.utexas.edu
%Date: Tue, 8 Jun 93 14:52:24 CDT

%%%%%%%%%%%%%%%%%%%%%%%%%%%%%%%%%%%%%%%%%%%%%%
% This paper requires amstex version 2.1.  If you do not have that, you can
% obtain it from the American Mathematical Society by telnetting to
% e-math.ams.com (IP number 130.44.1.100) and using the login and password
% ``e-math''. Instructions for ftp access can be obtained by following the
% instructions on the screen.  You will also need the amstex fonts.  If this
% fails, request a copy of the paper by sending email to me.
% (dafr@math.utexas.edu).
%
\input amstex
\input amsppt.sty
%%%%%%%%%%%%%%%%%%%%%%%%%%%%%%%%%%%%%%%%%%%%%%%

\CenteredTagsOnSplits
\NoBlackBoxes
\def\today{\ifcase\month\or
 January\or February\or March\or April\or May\or June\or
 July\or August\or September\or October\or November\or December\fi
 \space\number\day, \number\year}
\define\({\left(}
\define\){\right)}
\define\Ahat{{\hat A}}

\define\CC{{\Bbb C}}

\define\Diff{\operatorname{Diff}}

\define\End{\operatorname{End}}

\define\RR{{\Bbb R}}

\define\ZZ{{\Bbb Z}}
\define\[{\left[}
\define\]{\right]}

\define\chiup{\raise.5ex\hbox{$\chi$}}
\define\cir{S^1}

\define\exertag #1#2{\removelastskip\bigskip\medskip\eightpoint\noindent%
\hbox{\rm\ignorespaces#2\unskip} #1.\ }

\define\inv{^{-1}}
\define\mstrut{^{\vphantom{1*\prime y}}}
\define\protag#1 #2{#2\ #1}

\define\res#1{\negmedspace\bigm|_{#1}}
\define\temsquare{\raise3.5pt\hbox{\boxed{ }}}

\define\theprotag#1 #2{#2~#1}

\define\zmod#1{\ZZ/#1\ZZ}

%%%%%%%%%%%%%%%%%%%%%%%%%%
\def\bigstrut{\hbox{\vrule height18pt depth 9pt width0pt}}
\def\entry#1:#2:#3 {\bigstrut{#1}&{#2}&{#3}\cr}
\def\titlestrut{\hbox{\vrule height12pt depth 9pt width0pt}}
\define\CE{\Cal E}
\define\CV{\Cal V}
\define\CW{\Cal W}
\define\NatTrans{\operatorname{NatTrans}}
\define\Rep{\operatorname{Rep}}
\define\Sign{\operatorname{Sign}}
\define\TT{\Bbb T}
\define\Trace{\operatorname{Trace}}
\define\Z{$\ZZ$}
\define\bX{\partial X}
\define\bdct{\widetilde{{C}_{Y}}}
\define\bdc{C_{Y}}
\define\bo{\bold{1}}
\define\drig#1{\Diff^+_{\operatorname{rig}}(#1)}
\define\ger#1{\Cal{G}_{#1}}
\define\id{\operatorname{id}}
\define\mytimes{\odot}
\define\tger#1{\frak G_{#1}}
\define\tors#1{T_{#1}}
\define\tpi{2\pi i}
\define\zo{[0,1]}
%%%%%%%%%%%%%%%%%%%%%%%%%%%%%%%%%%%%%%%%%
%\magnification=\magstephalf
%\addto\tenpoint{\baselineskip=15pt}\tenpoint
\NoRunningHeads % USE IN FINAL VERSION; THEN COMMENT OUT NEXT LINE
\refstyle{A}
\widestnumber\key{RT}

	\topmatter
 \title\nofrills Extended Structures in Topological Quantum Field Theory
\endtitle
 \author Daniel S. Freed  \endauthor
 \thanks The author is supported by NSF grant DMS-8805684, a Presidential
Young Investigators award DMS-9057144, and by the O'Donnell
Foundation. This paper is based on a talk given in the {\sl Special Session
on Knots and Topological Quantum Field Theory\/} as part of the American
Mathematical Society meeting in Dayton, Ohio held in October, 1992.\endthanks
 \affil Department of Mathematics \\ University of Texas at Austin\endaffil
 \address Department of Mathematics, University of Texas, Austin, TX
78712\endaddress
%\curraddr \endcurraddr
 \email dafr\@math.utexas.edu \endemail
 \date June 8, 1993\enddate
%\dedicatory \enddedicatory
%\abstract \endabstract
	\endtopmatter

\document

\comment
\head
\S{}
\endhead
lasteqno @ 11
\endcomment

An $n$~dimensional quantum field theory typically deals with {\it partition
functions\/} and {\it correlation functions\/} of $n$~dimensional manifolds
and quantum Hilbert spaces of $n-1$~dimensional manifolds.  One of the novel
ideas in {\it topological\/} field theories is to extend these notions to
manifolds of dimension~$n-2$ and lower.  Such extensions inevitably lead to
the introduction of categories.  These ideas are very much ``in the air''.
Some of the people involved are Kazhdan, Segal, Lawrence, Kapranov,
Voevodsky, Crane, and Yetter.  Mostly this has been considered for
3~dimensional theories, but recently such ideas have also appeared in
relation to the 4~dimensional Donaldson invariants (see~\cite{Fu}, for
example).  Our motivation comes from a detailed understanding of {\it
classical\/} topological field theories, which we also extend to manifolds of
codimension two and higher.  In the particular case of gauge theory with
finite gauge group we define extensions of the usual ``path integral'' for
the extended classical theory~\cite{F1}.  For an $n$-manifold this is the
usual path integral, and for an $(n-1)$-manifold we recover the quantum
Hilbert space.  The result of this integration for an $(n-2)$-manifold is a
{\it 2-Hilbert space\/}.

In this note we briefly explain the consequences of this extended notion in
an arbitrary 3~dimensional topological theory.  The 2-Hilbert space~$\CE$ of
a circle has additional structure: it is a ``commutative, associative algebra
with identity and involution''.  This must be understood in the categorical
sense, since $\CE$~is a category.  In good cases $\CE$~can be realized as the
category of representations of a quantum group.  Hence we explain the
appearance of quantum groups in 3~dimensional field theories in terms of our
extended path integral.  We refer to~\cite{F1} for more details as well as
for an example: gauge theory with finite gauge group.

Reshetikhin/Turaev~\cite{RT} start with a Hopf algebra (of a certain type)
and from it construct a 3~dimensional field theory.  Recent work of
Kazhdan/Reshetikhin starts instead with a special type of category and
construct the field theory from it.  That category is then the 2-Hilbert
space of the circle in the resulting theory.  We find these algebraic data
unnatural in isolation; our purpose is to explain their introduction in terms
of general properties of field theory.  Our point of view is that they are
the ``solution'' to a field theory, rather than a natural starting point.  In
the last section we discuss framed tangles, and so make more direct contact
with the starting point in their work.  However, except in the case of finite
theories we cannot offer an alternative to their {\it constructions\/} of
invariants.

The 3~dimensional Chern-Simons theory with positive dimensional gauge
group~\cite{W} is only defined projectively for oriented manifolds; there are
central extensions of diffeomorphism groups which appear.  Witten realized
these central extensions by certain framings.  We briefly discuss a different
topological structure---a {\it rigging\/}---which realizes these central
extensions.  Our discussion is loosely based on some remarks in
Segal~\cite{S}.

 \subhead 2-Hilbert Spaces
 \endsubhead

A finite dimensional Hilbert space over~$\CC$ is a set~$W$ with operations of
addition, scalar multiplication, and hermitian inner product:
  $$ \aligned
     +\:W\times W&\longrightarrow W\\
     \cdot \:\CC\times W&\longrightarrow W\\
     (\cdot ,\cdot )\:W\times W&\longrightarrow \CC\endaligned  $$
These operations satisfy the usual axioms, which we do not list here.  A {\it
2-Hilbert space\/} has an analogous definition, except we replace the set~$W$
by a {\it category\/}~$\CW$ and the field~$\CC$ by the category~$\CV$ of all
finite dimensional Hilbert spaces.

A category differs from a set in that its ``elements'' (usually called
``objects'') may have automorphisms.\footnote{We abuse notation and
write~`$A\in \Cal{C}$' for an object~$A$ in a category~$\Cal{C}$.} More
generally, there are ``morphisms'' $V\to V'$ between objects in a category.
For example, the morphisms between any two inner product spaces $V,V'\in \CV$
are linear maps $V\to V'$.  But $\CV$~is much more than a category---it
has a structure analogous to a (semi)ring structure on a set.  Namely, there
is an addition (direct sum), a multiplication (tensor product), an additive
identity (the zero vector space), and a multiplicative identity (the ground
field~$\CC$).

So a 2-Hilbert space~$\CW$ is a module over~$\CV$ with an inner product.  In
other words, it is a category endowed with addition, scalar multiplication,
and an inner product:
  $$ \aligned
     +\:\CW\times \CW&\longrightarrow \CW\\
     \cdot \:\CV\times \CW&\longrightarrow \CW\\
     (\cdot ,\cdot )\:\CW\times \CW&\longrightarrow \CV\endaligned  $$
These maps are {\it functors\/}.  Of course, $\CV$~itself is a one
dimensional 2-Hilbert space; the inner product is
  $$ (V_1,V_2) = V_1\otimes \overline{V_2}.  $$
Analogous to the Hilbert space~$\CC^n$ is the $n$~dimensional 2-Hilbert
space~$\CV^n$ whose objects are $n$-tuples $\langle V^{(1)},\dots ,V^{(n)}
\rangle$ of inner product spaces.  Note that the multiplicative identity is
$\langle \CC,\dots ,\CC \rangle$; the inner product is
  $$ \bigl(\langle V^{(1)}_1,\dots ,V^{(n)}_1  \rangle\,,\, \langle
     V^{(1)}_2,\dots ,V^{(n)}_2  \rangle \bigr) = \bigoplus_{i=1}^n\,
     V^{(i)}_1\otimes \overline{V^{(i)}_2}.  $$

Since a 2-Hilbert space is a category, there is an extra layer of structure
beyond spaces and maps.  Namely, there are maps (morphisms) between elements
(objects) of a 2-Hilbert space, and so also maps (natural transformations)
between maps (functors).  Thus a familiar axiom for the hermitian inner
product now asserts the existence of a preferred isometry
  $$ (W_1,W_2) \longrightarrow \overline{(W_2,W_1)} \tag{1} $$
for any elements~$W_1,W_2$ in a 2-Hilbert space~$\CW$.  There are new
properties as well.  We assume the existence of preferred maps
  $$ \align
     C\longrightarrow (W,W) &\longrightarrow \CC \tag{2}\\
     (W_2,W_1)\cdot W_1 &\longrightarrow W_2 \tag{3}\endalign $$
for $W,W_1,W_2\in \CW$.  Note from~\thetag{1} that $(W,W)$~has a preferred
real structure, and we assume that \thetag{2}~is compatible.  The composition
is a real number attached to each~$W$, which we denote~$\dim W$.  The~`$\cdot
$' in~\thetag{3} is scalar multiplication.

Here is a less trivial example of a 2-Hilbert space.  Let $G$~be a finite
group and $\CW=\Rep(G)$ the category of finite dimensional unitary
representations of~$G$.  Addition is direct sum, scalar multiplication is
tensor product (with $G$~acting trivially on the ``scalar'' vector space),
and the inner product is
  $$ (W_1,W_2) = (W_1\otimes \overline{W_2})^G,  $$
the vector space of $G$-invariants in~$W_1\otimes \overline{W_2}$.  (There is
an additional operation---tensor products of representations---but we ignore
it for now.)  An {\it orthonormal basis\/} of~$\CW$ consists of a collection
of representations~$W_1,\dots ,W_n$, one from each equivalence class of
irreducible representations.  Then for any representation~$W$ the maps
in~\thetag{3} give a preferred isometry
  $$ \bigoplus_{i=1}^n\,\frac{1}{\dim W_i}\, (W,W_i)\cdot W_i\longrightarrow
     W. \tag{4} $$
Here $\frac{1}{\dim W_i}\, (W,W_i)$ is the vector space~$(W,W_i)$ with its
inner product multiplied by~$\frac{1}{\dim W_i}$.  This equation asserts that
$\{W_i/\sqrt{\dim W_i}\}$ is an ``orthonormal basis'' of~$\Rep(G)$.  If we
replace the finite group~$G$ by a compact Lie group of positive dimension,
then we obtain an infinite dimensional 2-Hilbert space.

Many constructions in linear algebra have analogues for 2-Hilbert spaces.
These include the dual space, direct sum, and tensor product.  The trace of
an endomorphism of a 2-Hilbert space is a Hilbert space.  For example, the
``dimension'' of~$\Rep(G)$ is the Hilbert space~$R(G)$ of equivalence classes
of representations of~$G$.  (There is also a ring structure on~$R(G)$ from
the tensor product of representations in~$\Rep(G)$.)

The reader may object that a 2-Hilbert space resembles an integral lattice in
a Hilbert space more than it does a Hilbert space.  Note, however, that
if~$V\in \CV$ is a Hilbert space, and $\mu \in \RR^+$ a positive real number,
then $\mu V\in \CV$ makes sense---it is the same underlying vector space~$V$
with inner product multiplied by~$\mu $.  Note also that $\mu V$~is isometric
to~$V$.  We can extend this multiplication formally to complex numbers with
nonzero phase, and then allow these scalars in 2-Hilbert spaces.  From this
point of view there is no natural lattice in~$\Rep(G)$, since we have nothing
to fix the scale of the inner product in an irreducible representation.  In
this regard there is an isometry~\thetag{4} with $W_i$~replaced by ~$\mu
_iW_i$ for {\it any\/} positive scalars~$\mu _i$.  The map in~\thetag{4} then
depends on the~$\mu _i$.

 \subhead Topological Field Theories
 \endsubhead

An $n$ dimensional topological field theory typically consists of assignments
  $$ \aligned
     Y^{n-1}&\longmapsto E(Y)\\
     X^n &\longmapsto Z_X\endaligned \tag{5} $$
of a finite dimensional Hilbert space~$E(Y)$ to a closed $(n-1)$-manifold,
and a ``path integral'' $Z_X\in E(\bX)$ to a compact $n$-manifold.  In most
theories the manifolds carry additional topological structure, such as an
orientation.  Symmetries of the manifolds which preserve the extra structure
are implemented as symmetries on the corresponding objects in the field
theory.  Thus symmetries of~$Y$ act as unitary transformations on~$E(Y)$ and
symmetries of~$X$ leave~$Z_X$ invariant.  Oppositely oriented manifolds map
to the conjugate object.  Most importantly, there is a gluing law for pasting
together components of the boundary of an $n$-manifold.  The axioms are
spelled out in~\cite{A1} (cf.~\cite{FQ}, for example).  They capture the
gross structure common to all topological theories; specific theories have
more detailed structure, of course.

The consequences of these axioms in a 2~dimensional topological field theory
are standard, and provide a good warmup to the 3~dimensional case.  Let
$E=E(\cir)$ denote the quantum Hilbert space of the circle.  Up to isotopy
there is only one orientation-reversing diffeomorphism of~$\cir$, and it
determines an {\it antilinear\/} map
  $$ c\:E\longrightarrow E. \tag{6} $$
Note that $c^2=\id$ since any orientation-preserving diffeomorphism of~$\cir
$ is isotopic to the identity.  So $c$~determines a {\it real structure\/}
on~$E$.  The path integral~$Z_X$ over any surface~$X$ is real.  The path
integral~$Z_P$ over the pair of pants~$P$ determines a multiplication
  $$ \circ \:E\otimes E\longrightarrow E,  $$
and $Z_{D^2}$~acts as the identity, where $D^2$~is the disk.  One easily
shows that the trilinear form
  $$ x\otimes y\otimes z\longmapsto \bigl(x\circ y,c(z) \bigr)_E,\qquad
     x,y,z\in E,  $$
is symmetric, and that the multiplication is associative.  It follows that
$E$~is semisimple.  Conversely, let $E$~be a commutative, associative
algebra~$E$ with identity, compatible real structure, and compatible inner
product.  Then we can construct a 2~dimensional field theory whose associated
algebra is~$E$.

We now extend the assignments~\thetag{5} to codimension two manifolds in an
$n$~dimensional theory.  As mentioned in the introduction, we understand this
extension in terms of a generalized ``path integral'' based on an extended
classical theory.  The result is that to a closed $(n-2)$-manifold~$S$ we
assign a 2-Hilbert space~$\CE(S)$:
  $$ S^{n-2}\longmapsto \CE(S).  $$
There are ``higher'' notions for lower dimensional manifolds, but they will
not concern us here.  The theory also assigns to an $(n-1)$-manifold~$Y$ with
boundary an element $E(Y)\in \CE(S)$.  We postulate a gluing law in terms of
the inner product on~$\CE(S)$ for $(n-1)$-manifolds pasted together
along~$S$.  We also assume that symmetries of~$S$ act on~$\CE(S)$ by unitary
transformations, and further that homotopies of symmetries act as natural
transformations (which respect the 2-Hilbert space structure.)  In
particular, this implies that $\pi _1\Diff(S)$ acts as unitary automorphisms
of the identity\footnote{If $\CE=\Rep(G)$ then a unitary automorphism of the
identity is multiplication by a phase on each irreducible representation.}
on~$\CE(S)$.  Compare with the standard assertion that for a closed
$(n-1)$-manifold~$Y$ there is an action of~$\pi _0\Diff(Y)$ by unitary
transformations of~$E(Y)$.

Our main interest is in 3~dimensional theories of oriented manifolds.  We
examine the consequences of the axioms for the structure of the 2-Hilbert
space~$\CE=\CE(\cir)$, where $\cir$~is the standard oriented circle.  (In the
next section we discuss some modifications necessary in a {\it projective\/}
theory, for example the Chern-Simons theory with positive dimensional compact
gauge group.)  First, $\pi _1\Diff^+(\cir)\cong \ZZ$ and the positive
generator, represented by a loop of rotations, acts as an automorphism of the
identity on~$\CE$.  That is, for each~$W\in \CE$ there is a morphism
  $$ \theta _W\:W\longrightarrow W \tag{7} $$
which commutes with all morphisms in~$\CE$.  Next, reflection induces a
conjugation $c\:\CE\to \CE$ as in~\thetag{6}, and we denote
  $$ c(W)=W^*. \tag{8} $$
Since~$c^2=\id$ we have $(W^*)^*=W$.  Then $\theta \mstrut _{W^*}=\theta
^*_W$ follows from an equation in~$\Diff(\cir)$ relating rotations and
reflections.

For any connected compact oriented 1-manifold~$S$ there is a unique isotopy
class of orientation-preserving diffeomorphisms $S\to\cir$.  In a
2~dimensional theory that suffices to identify the Hilbert spaces assigned
to~$S$ and~$\cir$ uniquely.  In a 3~dimensional theory, however, different
diffeomorphisms $S\to\cir$ yield different isometries $\CE(S)\cong
\CE(\cir)$, although there is an isometry (natural transformation) between
any two such isometries.  The latter depends nontrivially on a choice of
isotopy, due to the nontrivial~$\pi _1\Diff^+(\cir)$.  Hence we must make a
rigid convention to identify different circles.  Namely, we require any
circle~$S$ to appear in the complex line~$\CC$ and have its center
in~$\RR\subset \CC$.  Then there is a unique composition of a translation and
a dilation which identifies~$S$ with the standard circle~$\cir\subset \CC$.
The reflection which induces duality is reflection in the real axis.  We also
standardize surfaces diffeomorphic to a disk with a finite number of smaller
disjoint disks removed.  Such surfaces embed in~$\CC$ with standardized
boundaries.

Now $\CE=\CE(\cir)$ has a structure analogous to the real commutative
associative algebra structure (with compatible inner product) on~$E(\cir)$ in
a 2~dimensional theory.  The path integral~$Z_P$ on the standardized pair of
pants~$P$ determines a multiplication
  $$ \mytimes\:\CE\otimes \CE\longrightarrow \CE. \tag{9} $$
The ``reality'' statement is a preferred isometry
  $$ (W\mstrut _1\mytimes W\mstrut _2)^* \cong W_1^* \mytimes W_2^*
      $$
for all~$W_1,W_2\in \CE$.  The path integral~$Z_{D^2}=\bo\in \CE$ over the
standard disk~$D^2\subset \CC$ acts as an identity for the multiplication in
the sense that there are preferred isometries
  $$ \bo\mytimes W\cong W\mytimes \bo\cong W  $$
for all~$W\in \CE$.  The associativity is a natural isometry
  $$ \varphi _{W_1,W_2,W_3}\:(W_1\mytimes W_2)\mytimes W_3\longrightarrow
     W_1\mytimes (W_2\mytimes W_3) \tag{10} $$
which satisfies the ``pentagon relation''.  The isometry~\thetag{10} is
constructed from the gluing law, and the pentagon follows from cuttings and
pastings of the surface~$Q$ which is a disk with 3~interior disks removed.
Also, the braiding diffeomorphism $\beta \:P\to P$ of the pair of pants
induces a natural isometry
 $$ R_{W_1,W_2}\:W_1\mytimes W_2\longrightarrow W_2\mytimes W_1.  $$
Equations in~$\Diff^+(Q)$ imply two ``hexagon relations'', while the relation
  $$ (\theta \mstrut _{W_2}\mytimes\theta \mstrut _{W_1})\circ R\mstrut
     _{W_1,W_2} = R\inv _{W_2,W_1}\circ \theta \mstrut _{W_1\mytimes W_2}
      $$
follows from an equation in~$\Diff^+(P)$.

This is a taste of what may be extracted from the functoriality and the
gluing laws.  See~\cite{F1} for more details.  I imagine there is an
appropriate semisimplicity statement which can be made as well.

Category enthusiasts may prefer to regard~$\CE$ as an abelian category
endowed with extra structure---monoidal structure~\thetag{9}, duality or
rigidity~\thetag{8}, balancing~\thetag{7}, and braiding~\thetag{10}.  If we
are also given a {\it fiber functor\/}, that is a functor $\CE\to\CV$ which
preserves the monoidal structure, then $\CE$~can be recognized as the
category of representations~$\Rep(A)$ of a quasitriangular quasi-Hopf
algebra~$A$, also known as a quantum group.  In~\cite{F1} we construct an
obvious fiber functor for the Chern-Simons theory with finite gauge group,
and so recover Hopf algebras which appeared previously in this context.  In
general, a field theory does not construct a fiber functor, and they do not
exist for all theories.\footnote{For a finite $\sigma $-model into a space
with $n$~points, the 2-Hilbert space~$\CE$ is~$\CV^n$, which does not admit a
functor $\CV^n\to\CV$ which preserves direct sums and tensor products.
(Indeed, the image of the unit object~$\bo=\langle \CC,\dots ,\CC \rangle$
must be one dimensional, but $\bo$~is the sum of $n$~``basis'' elements.)
This is related to the fact that that the quantum Hilbert space of~$S^2$ is
$n$~dimensional.  In Chern-Simons theories, on the other hand, the Hilbert
space of~$S^2$ is one dimensional.  A related observation: The endomorphisms
of the unit object~$\bo=\langle \CC,\dots ,\CC \rangle$ form a ring which is
not a field.  I believe that an arbitrary unitary theory decomposes into a
direct sum of such ``irreducible'' theories---the idempotents in this ring
give such a decomposition---and it is possible that fiber functors always
exist for irreducible theories.}

We remark that if $\CE=\Rep(A)$ and $W\in \CE$ is an irreducible
representation, then $\theta _W$~is multiplication by~$e^{2\pi ih_W}$, where
$h_W$~is the {\it conformal weight\/} corresponding to~$W$.

Just as one can construct a 2~dimensional theory starting from an algebra~$E$
of the appropriate sort, so too one can construct a 3~dimensional theory
starting from a 2-Hilbert space~$\CE$ with multiplication, braiding, etc.  A
precise version of this statement is contained in recent work of
Kazhdan/Reshetikhin.

 \subhead Central Extensions
 \endsubhead

The Chern-Simons theory introduced by Witten~\cite{W} involves certain
central extensions, which he realizes in terms of ``2-framings''.  We follow
Segal~\cite{S} and instead propose that manifolds in the theory be endowed
with an extra topological structure termed a ``rigging''.  A rigging is a
trivialization of a topological invariant which for a closed oriented
4-manifold~$W$ is the signature~$\Sign(W)$.  There are corresponding
topological invariants of 3-, 2-, and 1-manifolds which we briefly explain
below.  Observe that three times the signature is the Pontrjagin
number~$p_1(W)\in \ZZ$.  Topological invariants in lower dimensions stemming
from~$p_1(W)$ are more easily constructed than those stemming
from~$\Sign(W)$.  The difference here is one between $K$-theory and
cohomology.  Also, there are invariants of spin manifolds which stem from the
$\Ahat$-genus~$\Ahat(W)$.  In all three cases we can define riggings, though
we focus here on the signature and associated invariants.  Our constructions
in this section are similar to constructions of Segal~\cite{S}.  This
material is preliminary as we cannot yet check all of the details.

A word about the abstractions which follow.  {\it Torsors\/} and {\it
gerbes\/} are concrete realizations of integral cohomology, somewhat
analogous to the way that elements of $K$-theory are realized by vector
bundles.  For example, a family of $\ZZ$-torsors is a principal $\ZZ$~bundle,
and it has a characteristic class in the first cohomology of the parameter
space.  Families of higher $\ZZ$-gerbes have characteristic classes in higher
integral cohomology.

Our starting point is the observation in~\cite{F2} that a Dirac operator on a
4-manifold~$W$ with boundary has a topological index which lives in a {\it
$\ZZ$-torsor\/} which is a topological invariant of the operator on the
boundary.  `$\ZZ$-torsor' is by definition `principal homogeneous space for
the integers'.  In the case of the signature operator that
$\ZZ$-torsor~$\tors{\partial W}$ has a canonical trivialization, and the
signature is, of course, an integer.  One can describe the
$\ZZ$-torsor~$\tors X(g)$ of a closed oriented 3-manifold~$X$ with metric~$g$
in terms of the $\xi $-invariant ($\frac 12\eta $-invariant) of
Atiyah/Patodi/Singer.  The $\xi $-invariant of the metric determines the
$\ZZ$-torsor~$\tors X(g)$ of real numbers~$x$ which satisfy $e^{\tpi
x}=e^{\tpi \xi }$.  The \Z-torsor~$\tors X$ is the space of sections of this
bundle of \Z-torsors over the space of metrics; it is a topological invariant
of~$X$.

Next, the exponentiated $\xi $-invariant of a Riemannian 3-manifold~$X$ with
boundary is meant to live in the determinant circle~$\bdc$ of the
boundary~$Y=\partial X$.  (We now omit the metric from the notation.)  This
is mentioned in~\cite{S} and is currently under investigation with Dai.  The
determinant circle is the set of elements of unit norm in the determinant
line with respect to the Quillen metric.  It is a torsor for the group~$\TT$
of unit size complex numbers.  Consider the collection of all
$\RR$-torsors~$\bdct$ which cover the $\TT$-torsor~$\bdc$, i.e., the
collection of all covering maps $\bdct\to\bdc$ compatible with the $\RR$~and
$\TT$~actions.  The set of morphisms between any two such is a \Z-torsor, and
the collection~$\ger{Y}$ of all these covering maps is an example of a {\it
\Z-gerbe\/}.  As before, we eliminate the choice of metric by working with
smooth families over the space of metrics.  This construction works for any
closed oriented 2-manifold~$Y$.  The exponentiated $\xi $-invariant of a
compact oriented 3-manifold is a point in~$C_{\partial X}$, and using this
point we can construct a particular cover~$T_X\in \ger{\partial X}$.

The corresponding topological invariant of a closed oriented 1-manifold~$S$
is more complicated to describe.  Briefly, for each metric on~$S$ there is a
self-adjoint signature operator, which is essentially two copies of the
operator~$i\frac{d}{dx}$ on functions.  It has discrete real spectrum
extending to both~$\infty $ and~$-\infty $.  Let $A\subset \RR$~be the
complement of the spectrum.  Then for $a,b\in A$ there is a finite
dimensional Hilbert space of eigenvectors with eigenvalue between~$a$
and~$b$.  Let $C_{a,b}$ be the determinant circle of this space.  These
circles fit together to form a flat circle bundle over~$A\times A$.  Now
consider a flat circle bundle~$C\to A$ together with consistent isomorphisms
$C_a\otimes C_{a,b}\to C_b$.  The collection of all such is a $\TT$-gerbe.
Then the collection of liftings of this $\TT$-gerbe to $\RR$-gerbes is a
``2-gerbe'' over~$\ZZ$.  Finally, we factor out the metric to obtain a
topological invariant~$\tger S$.

\midinsert
{\offinterlineskip \tabskip = 0pt
$$\vbox{  \halign{
	\vrule\enspace\hfil#\hfil\enspace\vrule\hskip2pt
        \vrule&\enspace\hfil#\hfil\enspace
	&\vrule\enspace\hfil#\hfil\enspace\vrule\cr
\noalign{\hrule}
\titlestrut{\bf dim}&{\bf closed manifold}&{\bf compact manifold with
boundary} \cr
\noalign{\hrule}
\vphantom{\vrule height 2pt}&&\cr \noalign{\hrule}
\entry 4:{$\Sign(W)\in \ZZ$}:{$\Sign(W)\in \tors{\bX}$}
\noalign{\hrule}
\entry 3:{$\vcenter{\vbox{\kern6pt\hsize120pt\centerline{$\ZZ$-torsor
$\tors X$}\vskip9pt\centerline{(lifts of $e^{\tpi\xi
}$)}\kern6pt}}$}:{$T_X\in \ger{\bX}$}
\noalign{\hrule}
\entry 2:{$\vcenter{\vbox{\kern6pt\hsize140pt\centerline{$\ZZ$-gerbe
$\ger Y$}\vskip9pt\centerline{(covers of determinant
circle)}\kern6pt}}$}:{$\ger Y\in \tger{\partial Y}$}
\noalign{\hrule}
\entry 1:{$\vcenter{\vbox{\kern6pt\hsize160pt\centerline{2 -gerbe
$\tger S$}\vskip9pt\centerline{(covers of ``determinant
$\TT$-gerbe'')}\kern6pt}}$}:{}
\vphantom{\vrule height2pt}&&\cr \noalign{\hrule}} }$$
}
\nobreak
\centerline{Table~1: Topological invariants of oriented manifolds}
\medskip
\endinsert

We summarize this discussion in Table 1. Each entry is a topological
invariant, which means it is functorial under orientation-preserving
diffeomorphisms.  The invariants also obey gluing laws, and they ``change
sign'' when the orientation is reversed.  As mentioned, the \Z-torsor~$\tors
X$ has a natural trivialization $\tors X\cong \ZZ$ when $X$~is closed.

A {\it rigging\/} of an oriented manifold is a trivialization of the
topological invariant in Table~1.  For 4-manifolds this is meaningless, or it
demands that we only consider 4-manifolds with vanishing signature.  For a
closed 3-manifold~$X$ a rigging is a choice of an element in~$\tors X$.
Recalling that $\tors X$~has a natural trivialization, this amounts to
choosing an integer.  The existence of a canonical element in~$\tors X$
corresponds to Atiyah's canonical 2-framing~\cite{A2}.  For a closed
2-manifold~$Y$ a rigging is a choice of cover of the determinant circle
bundle over the space of metrics.  I believe that our definition of a rigging
of a 1-manifold differs from Segal's.  Diffeomorphisms of rigged manifolds
are required to preserve the rigging.  Thus the group of
diffeomorphisms~$\drig Y$ of a closed oriented rigged 2-manifold~$Y$ is a
central extension by~$\ZZ$ of the group of orientation-preserving
diffeomorphisms~$\Diff^+(Y)$.  The fundamental group of rigged
diffeomorphisms~$\pi _1\drig S$ of a rigged oriented closed 1-manifold~$S$ is
a central extension by~$\ZZ$ of~$\pi _1\Diff^+(S)$.  Note that the boundary
of any oriented manifold is rigged.  Also, riggings glue together.

A 3~dimensional (unitary) field theory of rigged oriented manifolds has the
structure described in the previous section, but modified to account for the
riggings.  The example we have in mind here is Chern-Simons theory with
positive dimensional compact gauge group.  There are homomorphisms
$\ZZ\to\TT$ induced by: (i)\ the change in the path integral over a closed
3-manifold~$X$ under change of rigging; (ii)\ the action of the kernel of
$\pi _0\drig Y\to\pi _0\Diff^+(Y)$ on~$E(Y)$ for a closed rigged
2-manifold~$Y$; and (iii)\ the action of the kernel of $\pi _1\drig S\to\pi
_1\Diff^+(S)$ on~$\CE(S)$ (by automorphisms of the identity) for a closed
rigged 1-manifold~$S$.  I believe that these homomorphisms are universal in a
theory---that is, independent of~$X,Y,S$---and that the generator maps
to~$e^{\tpi c/24}$ where $c$~is the {\it central charge\/} of the theory.
Other computations must now take into account the riggings as well.  For
example, I believe that it is no longer necessarily true that the square of
the conjugation~\thetag{8} is the identity,\footnote{In Hopf algebra terms
this corresponds to asserting that the square of the antipode is not
necessarily the identity.} but I cannot yet see this in terms of riggings.

 \subhead Invariants of Framed Tangles
 \endsubhead

We briefly explain how a 3~dimensional topological field theory, extended as
previously discussed, yields invariants of framed tangles.  This is meant to
make contact with the Kazhdan/Reshetikhin work.\footnote{as explained to me
by Reshetikhin, who I warmly thank.}  See~\cite{RT} for a precise definition
of framed tangles.  For simplicity we ignore riggings in this section.

%\midinsert
%\bigskip
%\centerline{\epsffile{DIRECTORY/Figure1.ai}}
%\nobreak
%\centerline{Figure~1: A framed tangle}
%\bigskip
%\endinsert

A framed tangle~$D$ is represented by a diagram like Figure~1, where
$b$~strands intersect the line labeled~$0$ and $t$~strands intersect the line
labeled~$1$.  View the lines as copies of~$\CC$ and view the whole picture as
embedded in~$\RR^3$.  Extend the picture to~$\zo\times S^2$ by adding a point
at~$\infty $ to each plane and an extra strand~$\zo\times \{\infty \}$.  Now
cut out tubular neighborhoods of each strand to obtain a 3-manifold~$X$.  The
boundary of~$X$ is
  $$ \multline
      \partial X=\bigl(\{1\}\times S^2 \,-\,(t+1)D^2 \bigr)\;\cup\; -
     \bigl(\{0\}\times S^2 \,-\,(b+1)D^2 \bigr) \;\cup \;(\text{annulus at
     $\infty$} ) \\
      \hskip 80pt\cup\; \bigcup_{\pi _0(D)}(\text{annulus or torus} )
     .\hfill\endmultline  $$
Using the framings we can identify, up to isotopy, the annuli and tori with
the standard annulus $\zo\times \cir$ and the standard torus $\cir\times
\cir$. Now the quantum invariant~$Z_X$ is an element of the Hilbert
space~$E(\partial X)$.  We reinterpret it according to the decomposition
induced by~\thetag{4}.

Applying the gluing law to~\thetag{9} we find that $E(\{1\}\times
(S^2\,-\,(t+1)D^2))$ is the $(t-1)$-fold tensor product
  $$ \mytimes\:\CE\otimes \dots \otimes \CE\longrightarrow \CE. \tag{11} $$
We must choose a specific order for the multiplications since multiplication
is not associative (cf.~\thetag{10}).  For the annuli and tori we have
  $$ \aligned
     E(\zo\times \cir) &\cong (\id\:\CE\to\CE)\\
     E(\cir\times \cir) &= V\cong \Trace(\id\:\CE\to\CE).\endaligned
     $$
This last equation defines~$V$.  Now we glue the annuli to the
ends~$\{0\}\times S^2$ and~$\{1\}\times S^2$.  When we glue along a circle we
obtain an inner product in~$\CE$.  We only have to be careful about
orientations.  We geometrically pass between the two orientations of~$\cir$
via reflection, and so in the field theory via duality, according
to~\thetag{8}.

This abstract description can be made somewhat more concrete.  Notice that
by~\thetag{3} we can convert the inner products from the gluing into maps.
Write
  $$ \pi _0(D) = \pi _0(D)_{\text{annuli} }\;\cup\; \pi _0(D)_{\text{tori} }.
      $$
Define maps (functors)
  $$ F_i\:\underbrace{\CE\otimes \dots \otimes \CE}_{|\pi
     _0(D)_{\text{annuli} }|}\longrightarrow \CE,\qquad i=0,1,  $$
by
  $$ F_i(A_1\otimes \dots \otimes A_n) = A^{\pm}_{i_1}\mytimes \dots \mytimes
     A^{\pm}_{i_n},  $$
where $i_j$~is the position of the boundary of the $j^{\text{th} }$~annulus
among the circles in~$\{1\}\times S^2$, the sign is chosen according to the
orientation (a minus sign is the dual), and the product on the right hand
side is~\thetag{11}.  Then the vector space obtained by gluing the annuli to
the ends~$\{0\}\times S^2$ and~$\{1\}\times S^2$ maps to the vector space of
natural transformations from~$F_0$ to~$F_1$.  To include the tori we simply
take the tensor product with~$V^{\otimes |\pi _0(D)_{\text{tori} }|}$.  So,
finally, the partition function~$Z_X$ can be seen as an element of
  $$ \NatTrans(F_0,F_1)\otimes V^{\otimes |\pi _0(D)_{\text{tori} }|}.
      $$
This element is an invariant of the framed link.

This is almost the description of Kazhdan/Reshetikhin, except that they
realize~$V$ as the space of natural automorphisms of the trivial functor
$W\mapsto \bo$.  This is correct if~$\End(\bo)\cong \CC$.

%\midinsert
%\bigskip
%\centerline{\epsffile{DIRECTORY/Figure2.ai}}
%\nobreak
%\centerline{Figure~2: The natural transformation $W\mytimes W^*\to\bo$}
%\bigskip
%\endinsert

As a simple example we see that the partition function of the tangle
pictured in Figure~2 is a natural transformation $W\mytimes W^*\mapsto\bo$.
Similarly, we find a natural transformation $\bo\to W\mytimes W^*$.  This
fills a gap in~\cite{F1,\S5}.

%\newpage


\Refs

\ref
\key A1
\by M. F. Atiyah
\paper Topological quantum field theory
\jour Publ. Math. Inst. Hautes Etudes Sci. (Paris)
\vol 68
\yr 1989
\pages 175--186
\endref

\ref
\key A2
\by M. F. Atiyah
\paper On framings of 3-manifolds
\vol 29
\pages 1--7
\yr 1990
\jour Topology
\endref

\ref
\key F1
\by D. S. Freed
\paper Higher algebraic structures and quantization
\jour Commun. Math. Phys.
\toappear
\endref

\ref
\key F2
\by D. S. Freed
\paper A gluing law for the index of Dirac operators
\miscnote to appear in the 60th birthday volume dedicated to Richard Palais
\endref

\ref
\key FQ
\by D. S. Freed, F. Quinn
\paper Chern-Simons theory with finite gauge group
\jour Commun. Math. Phys.
\toappear
\endref

\ref
\key Fu
\by K. Fukaya
\paper Floer homology for 3-manifolds with boundary
\miscnote University of Tokyo preprint, 1993
\endref

\ref
\key RT
\by N. Reshetikhin, V. G. Turaev
\paper Invariants of 3-manifolds via link polynomials and quantum groups
\jour Invent. math.
\yr 1991
\vol 103
\pages 547--597
\endref

\ref
\key S
\by G. Segal
\paper The definition of conformal field theory
\miscnote preprint
\endref

\ref
\key W
\by E. Witten
\paper Quantum field theory and the Jones polynomial
\jour Commun. Math. Phys.
\vol 121
\yr 1989
\page 351--399
\endref

\endRefs
\enddocument
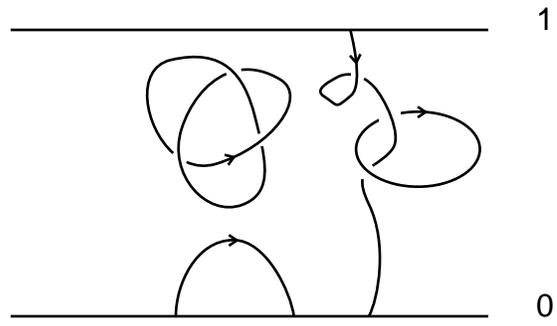
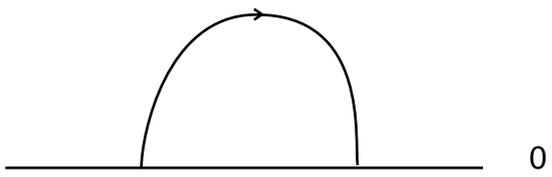
